\documentclass[aps,prb,reprint,twocolumn,superscriptaddress,showpacs,floatfix]{revtex4-1}

\usepackage{hyperref}
\usepackage{graphicx}
\usepackage{amsmath}
\usepackage{amssymb}
\usepackage{bbm}
\usepackage{color}
\usepackage{enumitem}

\newcommand{\Tr}{\text{Tr}}

\begin{document}

\title{Finite temperature quantum embedding theories for correlated systems}
\author{Dominika Zgid}
\email{zgid@umich.edu}
\thanks{Corresponding author}
\affiliation{Department of Chemistry, University of Michigan, Ann Arbor, Michigan 48109, USA}
\author{Emanuel Gull}
\affiliation{Department of Physics, University of Michigan, Ann Arbor, Michigan 48109, USA}

\begin{abstract}
The cost of the exact solution of the many-electron problem is believed to be exponential in the number of degrees of freedom, necessitating approximations that are controlled and accurate but numerically tractable. In this paper, we show that one of these approximations, the self-energy embedding theory (SEET), is derivable from a universal functional and therefore implicitly satisfies conservation laws and thermodynamic consistency. We also show how other approximations, such as the dynamical mean field theory (DMFT) and its combinations with many-body perturbation theory, can be understood as a special case of SEET and discuss how the additional freedom present in SEET can be used to obtain systematic convergence of results.
\end{abstract}
\maketitle
\section{Introduction}
The computational cost of the exact solution of the realistic extended many-electron problem is believed to be exponential in the number of degrees of freedom, necessitating the development of accurate approximate methods able to capture interacting electron physics.\cite{Dirac29}

While mature tools for obtaining ground state energetics for both molecular and solid state problems exist,\cite{Kohn65,MartinInteracting16} solid state  experiments are often performed at finite temperature and yield as the measured result not energy differences but single-and two-particle response functions, requiring a description of finite temperature excitations.

Many-body perturbation theory\cite{MartinInteracting16} accurately describes these phenomena where interactions are weak. However, many systems of interest are believed to be outside the regime of validity of perturbative approximations. In these systems, a non-perturbative solution is desired for a subset of the correlated degrees of freedom embedded into a background of more weakly correlated, perturbatively treated states.
Ideally such an embedding construct should be numerically tractable and defined in terms of  one or more small parameters that allow its tuning from a crude but computationally cheap, approximate solution to the exact but exponentially expensive one.

Several such theories have been developed. They include the dynamical mean field theory (DMFT),\cite{Georges96,Kotliar06} its combination with electronic structure methods, such as LDA+DMFT~\cite{Anisimov97,Lichtenstein98,Sun02}  and GW+DMFT~\cite{Biermann03,Biermann05}, the self-energy functional theory,\cite{Potthoff03} and most recently the self-energy embedding theory (SEET).\cite{Zgid15,Tran15b,Tran16} All of them require a compromise between accuracy and numerical tractability or time to solution.

In this paper, we show that SEET can be understood as a conserving functional approximation to an exact Luttinger-Ward functional.\cite{Luttinger60} This functional framework of SEET allows us to compare this theory to other functional approximations, and show in particular that DMFT, HF+DMFT, and GW+DMFT can be understood as a special case of SEET and to illustrate how the additional freedom given by SEET can be employed to systematically improve results.
In particular, we focus on various aspects of electron `screening' and downfolding  and how they are treated in various approximations.

This paper proceeds as follows. In Sec.~\ref{sec:System}, we introduce the system under study, the SEET definition, DMFT, and  several combinations of DMFT with many-body perturbation theory. In Sec.~\ref{sec:relationship}, we compare the different approaches based on their functionals. In Sec.~\ref{sec:Screening}, we focus in detail on various aspects of electron screening. We form conclusions in Sec.~\ref{sec:Conclusions}.

\section{System and formalism}\label{sec:System}
We consider a system described by a Hamiltonian with full two-body interaction $v_{ijkl}$ and one-body terms $t_{ij}$ in a finite orbital basis:
\begin{align}
H=\sum_{ij}^Nt_{ij}a^{\dagger}_{i}a_{j}+\sum_{ijkl}^Nv_{ijkl}a^{\dagger}_{i}a^{\dagger}_{j}a_{l}a_{k}, \label{eqn:realistic_ham1}
\end{align}
where the indices $i$, $j$, $k$, and $l$ enumerate all $N$ basis orbitals present in the system. In case of a periodic system,  Eq.~\ref{eqn:realistic_ham1} may in particular contain one-body terms connecting any orbital in any unit cell to any other orbital in any other unit cell, and general two-body integrals $v$ mixing interactions between any of the orbitals in any of the unit cells in the system.

Physical properties including thermodynamic quantities (energies and entropies), frequency dependent single-particle (Green's functions and self-energies) and two-particle quantities (susceptibilities) can be described in a functional approach. \cite{Luttinger60,Baym62,Albladh99,Potthoff06} In this approach, a $\Phi$- functional $\Phi[G]$ of the Green's function $G$, which contains all linked closed skeleton diagrams,\cite{Luttinger60} is used to express the grand potential as
\begin{align}
\Omega = \Phi - \Tr \log G^{-1} - \Tr \Sigma G,
\end{align}
and it satisfies
\begin{align}
\frac{\delta \Phi}{\delta G}=\Sigma[G], \label{eq:dphidg}
\end{align}
where the self-energy $\Sigma$ is defined with respect to a non-interacting Green's function $G_0$ via the Dyson equation
\begin{align}
G=G_0 + G_0 \Sigma G.\label{eq:dyson}
\end{align}
The functional formalism is useful because approximations to $\Phi$ that can be formulated as a subset of the terms of the exact $\Phi$ functional can be shown to respect the conservation laws of electron number, energy, momentum, and angular momentum by construction.\cite{Baym61,Baym62} In addition, $\Phi$-derivability ensures that quantities obtained by thermodynamic or coupling constant integration from non-interacting limits are consistent.\cite{Baym62} Functional theory therefore provides a convenient framework for constructing perturbative \cite{Baym62,Hedin65,Bickers89,Bickers91} and non-perturbative \cite{Georges96,Potthoff03,Zgid15,Tran15b,Tran16} diagrammatic approximations.

On the other hand, approximations based on a $\Phi$ functional do not guarantee self-consistency on the two-particle level, so that vertex functions which appear in the calculation of the one-particle self-energy may not the same as those generated by functional differentiation in two-particle correlation functions, and crossing symmetries may be violated.\cite{Bickers04,DeDominicis64,DeDominicis64B} The construction of methods for model systems that respect these symmetries by construction is an active topic of research.\cite{Rohringer16,vanLoon16}

The approximations we discuss in the following sections 
are all expressed in the functional form, thus making them straightforward to discuss and compare their respective assumptions, limits, and strengths.

\subsection{The Self-energy Embedding Theory}
\subsubsection{Self-energy Embedding Equations}
The self-energy embedding theory (SEET) \cite{Zgid15,Tran15b,Tran16} starts from the assumption that all orbitals present in the system can be separated into $M$ distinct orbital subsets $A_i$,  each containing $N^A_i$ orbitals, and a remainder $R$ with $N^R$ orbitals, such that $N^A_i \ll N$, for each $i$, and $N=\sum_{i=1}^M N_i^A+N^R$.

We assume that the orbitals within each subset $A_i$ are more strongly correlated among each other than with other orbitals present in the system, so that their intra-subset correlations need to be obtained in a non-perturbative way.
Conversely, inter-set correlations between orbitals belonging to two different sets $A_i$ and $A_j$, $i\neq j$, and correlations belonging to the remainder $R$
are assumed to be weaker, such that they can be simulated perturbatively. 
The choice of orbital subsets and subset size $N^A_i$ is general and will be commented on in Sec.~\ref{sec:orbitalselec}.

SEET first approximates the solution of the entire system using an affordable but potentially inaccurate $\Phi$-derivable method (weak coupling methods are a natural choice), and then corrects this approximation in the strongly correlated subspaces by a non-perturbative result. This is achieved by approximating the exact $\Phi$-functional as
\begin{align}\label{eq:SeetPhi}
\Phi^\text{SEET} = \Phi^\text{tot}_\text{weak} + \sum_{i=1}^{M} \Big([\Phi_\text{strong}^A]_i-[\Phi_\text{weak}^A]_i\Big).
\end{align}
Here, $\Phi^\text{tot}_\text{weak}$ denotes a solution of the entire system using a conserving low-order approximation, for instance self-consistent second order perturbation theory (GF2)\cite{Dahlen05,Zgid14,Rusakov16,Phillips15,Kananenka15,Kananenka16} or the GW method.\cite{Hedin65}
$\Phi^A$ denotes all those terms in $\Phi$ where all  four indices $i,j,k,l$  of $v_{ijkl}$ are contained inside orbital subspace $A$.
$\Phi^A_\text{weak}$ is the approximation to $\Phi^A$ within the weak coupling method used for solving the entire system, and $\Phi^A_\text{strong}$ the approximation or exact solution of $\Phi^A$ obtained using the higher order method capable of describing `strong correlation'. 

Since the self-energy is a functional derivative of the $\Phi^\text{SEET}$-functional,  
 the total self-energy $\Sigma$ contains diagrams from both the `strong' and `weak' coupling methods and can be written in a matrix form reflecting the system separation onto different correlated blocks
 \begin{eqnarray}\label{eqn:sigma_seet}
 \Sigma^\text{SEET}=
 \begin{bmatrix}
    [\Sigma^\text{A}]_{1} & \Sigma^\text{int} & \dots &\dots &\dots\\
       \Sigma^\text{int}  &  [\Sigma^\text{A}]_{2} & \Sigma^\text{int} & \dots &\dots \\
     \dots & \dots &  \dots &  \dots &\dots\\
      \dots &  \dots &\Sigma^\text{int} & [\Sigma^\text{A}]_{M} &  \Sigma^\text{int} \\
  
   \dots &  \dots & \dots &    \Sigma^\text{int} & \Sigma^{R}
 \end{bmatrix}
 \end{eqnarray}
These blocks are obtained upon differentiation of the $\Phi^{SEET}$ functional according to Eq.~\ref{eq:dphidg} and have the following form
\begin{align}
[\Sigma^A]_{i}&=\Sigma^\text{tot}_\text{weak}+([\Sigma^{A}_\text{strong}]_{i}-[\Sigma^{A}_\text{weak}]_i),\label{eq:seet}\\
\Sigma^R&=\Sigma^{R}_\text{weak},\\
\Sigma^{int}&=\Sigma^{int}_\text{weak}.
\end{align}
Eq.~\ref{eqn:sigma_seet} describes a subspace self-energy consisting of a contribution from the strongly correlated subspace embedded into a  weakly correlated self-energy generated by all orbitals outside the subspace. This embedding of the  self-energy leads to the name `self-energy embedding theory'.

SEET satisfies the following limits:
\begin{itemize}
\item If the interaction $v_{ijkl}$ is zero or the temperature is infinity, $\beta=0$, the self-energy is zero and therefore the method becomes exact.
\item If $M=1$ and the only subspace $A$ includes all orbitals present in the system, $N^A=N$, so that no orbitals are left in the perturbatively treated subspace, $N^R=0$, then the entire system is solved using the strong correlation method and $\Phi^{SEET}=\Phi^{A}_\text{strong}$. Consequently, if the strong correlation method provides the exact solution, the exact solution of Eq.~\ref{eqn:realistic_ham1} is recovered.
\item In the limit of non-interacting subsystems, when the interactions between strongly correlated subspaces are zero, together with a condition $N^R=0$ and $\sum_i N^{A}_i=N$,  SEET recovers the solution of the system with the strong correlation method since $\Phi^\text{SEET}=\sum_{i}^{M}[\Phi^{A}_\text{strong}]_i$.
\item If the correlated subspaces are not treated exactly but using the same `weak correlation' method as the rest of the system, the weak correlation solution for the full system is recovered since $\Phi^\text{tot}_\text{weak} = \Phi^\text{tot}_\text{weak} + \sum_{i=1}^{M} \Big([\Phi_\text{weak}^A]_i-[\Phi_\text{weak}^A]_i\Big)$.
\end{itemize}

While consideration of the exact limits is essential, the important  practical question is whether (and where) one can expect  SEET to be accurate away from these exact limits. As is evident in Eq.~\ref{eq:SeetPhi}, SEET becomes accurate where the diagrams considered at the lower level method require no higher order corrections. This is the case in the high temperature, high energy, and high doping regimes where the self-energy is perturbative. Additionally, SEET is accurate if all non-perturbative correlations are restricted to the correlated subspaces, and its accuracy will therefore strongly depend on the choice of the correlated subspaces. 
Consequently, choosing the correlated subspaces is an important step in any SEET calculation.

\subsubsection{The choice of SEET subspaces}\label{sec:orbitalselec}
\begin{figure} [htb]
  \includegraphics[width=\columnwidth]{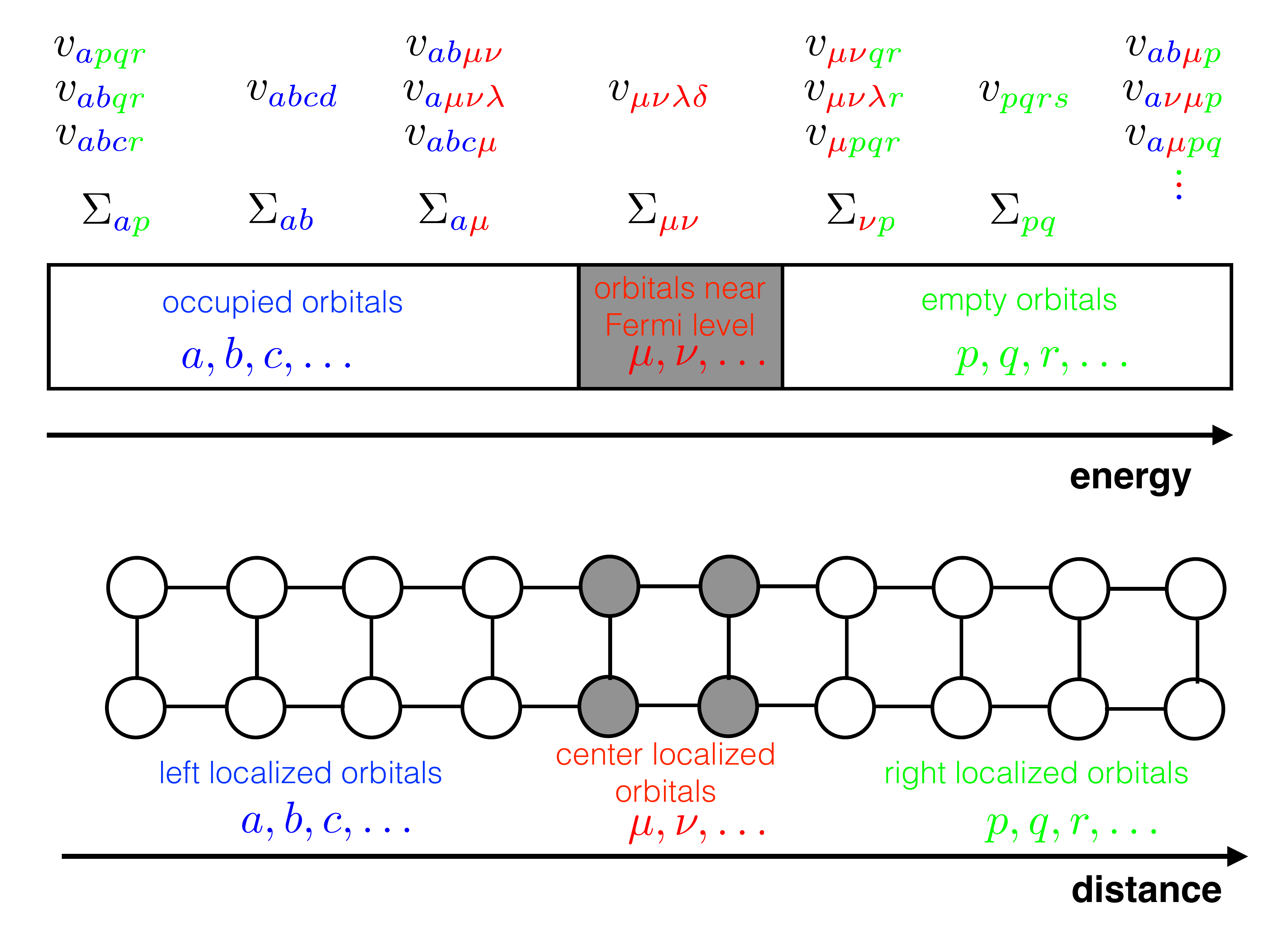} 
  \caption{Illustration of two choices of SEET subspaces. Top panel: Selection of orbital subspaces based on the energy/occupation scheme: partially occupied orbitals near the Fermi level ($\mu,\nu$) are included in the correlated subspace, any other contribution is excluded. Bottom panel: Selection of orbitals based on a localization criterion: sets of neighboring orbitals are chosen as the correlated subspace.}
  \label{fig:seet_separation}
\end{figure}
In many techniques, the `strongly correlated' and `weakly correlated' orbital subsets are chosen a priori. An example is DMFT, where $\Phi$ is truncated to local degrees of freedom,\cite{Kotliar06} or LDA+DMFT methods\cite{Anisimov97,Georges04,Kotliar06} where certain `local' orbitals (usually orbitals with $d$ or $f$-like character) are considered to be `correlated', while `wider' $p$ and $s$ orbitals are considered to be non-interacting.

While the same ad hoc orbital choice can be used for self-energy embedding theory, SEET also offers a different approach to the selection of correlated orbitals and in particular makes an adaptive choice of correlated orbitals `a posteriori' possible, without the need to localize or `downfold' orbitals.

A simple criterion for identifying the degree of orbital correlation is given by the frequency-dependence of the self-energy: the larger the frequency dependent part, the more `non-Hartree-Fock' like an orbital is, and therefore the more it needs to be treated at the `strongly correlated' level. 
Since Hartree-Fock only yields orbital occupancies of 0 and 2 (at zero $T$), then any partial occupancy of an orbital obtained from diagonalizing a one-body density matrix obtained using a perturbative approach (used in the first step of SEET) indicates some degree of correlation.
The larger the deviation from 0 or 2 the more ``strongly correlated'' an orbital is and the more likely it requires a non-perturbative treatment.

Consequently, the SEET calculations in  Ref.~\onlinecite{Zgid15,Tran15b,Tran16} added orbitals to the strongly correlated subspace $A$ using a criterion based on diagonalization of the one-body density matrix: chosen were those $N_A$ orbitals with the largest deviation of the occupancy from $0$ and $2$. This requires a basis transform of the hybridization function, non-interacting Hamiltonian, and two-body integrals into the basis that diagonalizes the one-body density matrix. While basis transforms for the two-body integrals are generally expensive, the transformed integrals are only necessary inside the correlated subspace, making the transform affordable in practice, such that the orbital transformation step is not a computational bottleneck.

Two possible subset selection schemes are illustrated graphically in Fig.~\ref{fig:seet_separation}. The upper panel shows a separation of orbitals based on energy or occupation scheme, where mostly unoccupied and mostly filled orbitals are treated as weakly correlated subspaces that can be treated by a weak correlation method. Partially filled orbitals are chosen as strongly correlated that will be treated by a non-perturbative method. The lower panel shows an alternative separation based on distance, where orbitals localized around a center position are treated as strongly correlated, whereas orbitals at farther distance are treated as uncorrelated or weakly correlated.

\subsubsection{Self-consistent solution of the SEET equations}
The $\Phi^\text{SEET}$ functional of Eq.~\ref{eq:SeetPhi} defines the  SEET approximation.\footnote{Potentially multiple self-consistent solutions exist, in analogy to Ref.~\onlinecite{Kozik15,Rusakov16,Welden16}.} It requires the specification of the $M$ correlated orbital subspaces $A_i$ and the subspace $R$, in addition to  the `strong coupling' and the perturbative weak coupling diagrams. We now describe an algorithm that generates a self-consistent solution of the SEET equations.

First, the weak coupling method is used to self-consistently obtain the self-energy $\Sigma^\text{tot}_\text{weak}$ and functional $\Phi^\text{tot}_\text{weak}$ of the entire system from a given initial Green's function, {\it e.g.} the Hartree-Fock (HF) or density functional theory (DFT) approximation. The self-consistency of the weakly correlated method eliminates all memory of the initial starting point in its convergence to a fixed point.
Upon convergence of the weakly correlated method, we choose the correlated subspaces according to Sec.~\ref{sec:orbitalselec}. We then compute $[\Sigma_\text{weak}^{A}]_i$ and $[\Phi_\text{weak}^{A}]_i$ in every orbital subspace $i$, {\it i.e.} the weak correlation approximation obtained with vertex indices exclusively contained in the correlated orbital subsets $A_i$.

In a next step, $[\Sigma_\text{strong}^A]_i$ needs to be obtained in each subspace $i$. To simplify notation, we select one particular subspace $A_i=A$ and absorb all other subspaces $A_j, j\neq i$, and the remaining weakly correlated orbitals in space $R$.
Using the non-interacting Green's function~\footnote{Note that there is a freedom of choice for the non-interacting Green's function. While we are using $G_0=(\omega-t)^{-1}$ here, where $t$ is the kinetic plus nuclear-electron attraction part of the Hamiltonian, in general other definitions of the one-body Hamiltonian  are possible. One of the most commonly used definitions for realistic systems is $t=F$, where $F$ is a Fock matrix obtained from HF, GF2, or GW.} 
in a block form
\begin{align}
G_0=\begin{pmatrix}\omega - t_{A}& -t_\text{int}\\
-t^{\dagger}_\text{int}& \omega - t_{R}\end{pmatrix}^{-1} \label{eq:g0eq}
\end{align}
and the Dyson equation $G = G_0+ G_0\Sigma G,$ we express the interacting Green's function as
\begin{align}\label{Gweak}
G^\text{tot}=\begin{pmatrix}(G_0^{-1})^{A}-\Sigma^{A}& (G_0^{-1})^\text{int}-\Sigma^\text{int}\\
\left[(G_0^{-1})^\text{int}-\Sigma^\text{int}\right]^{\dagger} &(G_0^{-1})^{R}-\Sigma^{R}\end{pmatrix}^{-1},
\end{align}
where $(G_0^{-1})^{A}$ denotes the inverse of the non-interacting Green's function restricted to the orbital subset $A$. Evaluation of $G^\text{tot}$
in the subset $A$ yields
\begin{align}
(G^\text{tot})^{A} &= \Big( (G_0^{-1})^{A} - \Sigma^{A} - \Delta \Big)^{-1}, \label{eq:subsetprop}
\end{align}
where $\Delta$ is defined as
\begin{align}
\Delta &= 
\Big[\left[(G_0^{-1})^\text{int}-\Sigma^\text{int}\right]^{\dagger} \times \label{eq:hybfun}\\ \nonumber & \left[(G_0^{-1})^{R}-\Sigma^{R}\right]^{-1}\left[(G_0^{-1})^\text{int}-\Sigma^\text{int}\right]\Big].
\end{align}
Eq.~\ref{eq:SeetPhi}, Eq.~\ref{eq:subsetprop} and Eq.~\ref{eq:hybfun} show that the `strongly correlated' $A$-subspace problem can be entirely formulated in the strongly correlated subspace as a problem in which the original interactions $v_{ijkl}$ have been restricted to the subspace $A$, but for which the bare Green's functions have been modified from $G_0$ to new propagators $\mathcal{G}_0$ which contain a contribution from a frequency-dependent `hybridization function' $\Delta$. These propagators are defined as
\begin{align}\label{eqn:g0}
\mathcal{G}_0^{-1}= (G_0^{-1})^{A} - \Delta.
\end{align}
Problems of this type are known as quantum impurity problems.  A quantum impurity solver will obtain an expression for a correlated $(G^\text{imp})^A$ given $\Delta$ (Eq.~\ref{eq:hybfun}) and $G_0$ (Eq.~\ref{eq:g0eq}) as well as a subset of interactions $v_{ijkl} \in A_i$ in either spatial or energy basis. Using the impurity problem Dyson equation, the self-energy for a strongly correlated orbital subset is obtained as  
\begin{equation}\label{eqn:dyson_imp}
[\Sigma^A_\text{strong}]_i=\mathcal{G}_0^{-1}-((G^\text{imp})^A)^{-1}.
\end{equation}
Once this strongly correlated $\Sigma_\text{strong}^A$ is known, the total self-energy, $\Sigma^A$, in subspace $i$ is evaluated as
\begin{align}\label{eqn:sigma_tot}
[\Sigma^A]_{i}&=\Sigma^\text{tot}_\text{weak}+([\Sigma^{A}_\text{strong}]_{i}-[\Sigma^{A}_\text{weak}]_i).
\end{align}
We note in particular that there are contributions to the $A$-subspace self-energy from vertices and propagators with some indices outside of subspace $A$. These contributions are contained within $(\Sigma^\text{tot}_\text{weak}-[\Sigma^{A}_\text{weak}]_i)$ and only treated at the perturbative level. We would like to stress that these contributions provide an effective adjustment caused by non-local interactions to the $[\Sigma^{A}_\text{strong}]_{i}$ that was evaluated using a subset of local interactions $v_{ijkl} \in A_i$.

While quantum impurity models were originally formulated in the context of dilute impurities in a metal,\cite{Anderson61} they form the basis of many non-perturbative embedding schemes including DMFT.\cite{Georges96,Kotliar06} Impurity problems are numerically tractable, with accurate or numerically exact methods ranging from continuous-time quantum Monte Carlo\cite{Rubtsov05,Werner06,Werner06b,Gull08,Gull11} to exact diagonalization \cite{Caffarel94,ED_Liebsch}, configuration-interaction \cite{Zgid12}, and numerical renormalization group theory \cite{Bulla08} methods. The requirements for SEET impurity problems, {\it i.e.} general (`non-diagonal') hybridization functions $\Delta$, multiple impurity and bath orbitals, and general interactions $v_{ijkl}$ currently make methods based on the configuration interaction hierarchy\cite{Zgid11, Zgid12} most suitable for this task, despite the necessity to approximate the continuous hybridization function $\Delta$ by a set of discrete bath levels and bath couplings.

If multiple correlated spaces are present, separate impurity problems need to be solved in each subspace $A_i$, and correlated self-energies $[\Sigma_{A}]_i$ obtained. These self-energies are then used to update each $[\Sigma_{A}]_i$ block of the self-energy $\Sigma^\text{SEET}$ obtained with the weak coupling method according to Eq.~\ref{eqn:sigma_seet}, and the Green's function for the entire system is evaluated using the Dyson equation. Iteration of this procedure, alternating weak coupling steps to update $\Sigma^{int}, \Sigma^{R},$ with impurity solver steps to obtain $[\Sigma^{A}]_i$ produces a converged $\Phi^\text{SEET}$ and $\Sigma^\text{SEET}$ of the form of Eq.~\ref{eq:SeetPhi}.  Appendix~\ref{app:SEETAlgo} and Refs.~\onlinecite{Zgid15, Tran15b,Tran16} have detailed step-by-step instructions on the construction of the iterative procedure.

\section{Relationship to other functional based theories}\label{sec:relationship}
\subsection{DMFT}
DMFT~\cite{Metzner89,Georges92,Georges96} is a $\Phi$-derivable theory that can be cast as an approximation to the exact $\Phi$ functional \cite{Kotliar06}:
\begin{align}
\Phi_\text{DMFT} = \sum_{j=1}^M [\Phi_I]_j\label{eq:phidmft}
\end{align}
where $j$ denotes unit cells, and $[\Phi_I]_j$ contains all those diagrams of $\Phi$ where the interaction vertices have all four indices inside unit cell $j$. All diagrams in $\Phi$ connecting different unit cells, either via interactions or via propagators, are discarded.
As a consequence, $\Sigma_\text{DMFT} = \frac{\delta \Phi_\text{DMFT}}{\delta G}$ is purely local to every cell.
In a translationally invariant system where all unit cells are equal, $\Sigma_\text{DMFT}$ is independent of $I$, and only one impurity problem exists.
In analogy to Eq.~\ref{eq:subsetprop}, an impurity model with $G^\text{imp}=G_I$, $\Sigma^\text{imp}=\Sigma_I$ can be defined and the self-consistent solution of the Dyson equation $G=G_0 + G_0 \Sigma_\text{DMFT}G$ and the solution of the impurity problem leads to the DMFT approximation of Eq.~\ref{eqn:realistic_ham1}.\footnote{Note that it is also possible to consider DMFT as an approximation to the momentum conservation at the vertices, where all terms of $\Phi$ are considered but the propagators are replaced with local propagators.\cite{Hettler00} This violates momentum conservation at each vertex.}

Eq.~\ref{eq:phidmft} shows that DMFT can be understood as a special case of SEET in which the orbital subspaces $A_i$ are chosen to be the orbitals local to a unit cell, the `weak correlation' method is skipped so that $\Phi_\text{weak}=0$, and the strong-correlation problem is computed by the DMFT impurity solver. Correspondingly, DMFT will provide a good approximation to the physics of a correlated system as long as the following two criteria are fulfilled: first, the interactions are predominantly local; and second, self-energy contributions from non-local terms (interactions or propagators) are negligible.

\subsection{HF+DMFT}
Similarly, HF+DMFT can be cast into this framework. The chosen correlated orbital subspaces $I_j$ are local to each unit cell, and the exact $\Phi$ is approximated as
\begin{align}
\Phi_\text{DMFT+HF} = \Phi^\text{tot}_\text{HF}+\sum_{j=1}^{M} \left([\Phi^I]_j -  [\Phi_\text{HF}^I]_j\right) \label{eq:phidmfthf},
\end{align}
where $[\Phi_\text{HF}^I]_j$ is the HF $\Phi$-functional with vertex indices restricted to unit cell $j$.
To obtain a self-consistent $\Phi_\text{DMFT+HF}$, the Hartree Fock equations are solved for the entire system and subsequently some or all local orbitals are chosen to the correlated subspace $I_j$. The impurity problem is then solved in the local subspace along the lines of DMFT. 

Note that all the non-local contributions to the self-energy of the unit cells that are frequency independent are generated by $\Phi^\text{tot}_\text{HF}$.
Any higher order contributions to the self-energy that are frequency dependent have purely local vertices and there are no non-local frequency dependent self-energy terms in the $\Phi_\text{DMFT+HF}$ functional.  Additionally, in the non-empirically adjusted HF+DMFT all impurity interactions remain the bare Coulomb interactions $v_{pqrs}$ and are local to the unit cell orbital subspaces $I_j$.

Consequently, any adjustment or renormalization of the frequency dependent   $[\Sigma^{A}_\text{strong}]_i$ term due to the non-local effects that is present in SEET(ED-in-GF2 or ED-in-GW) is absent in HF+DMFT. This is the reason why spectral features and energies produced at the HF+DMFT or LDA+DMFT level using a bare, unrenormalized local Coulomb interaction are not recovered correctly. For small molecular systems, the incorrect energies resulting from employing HF+DMFT with bare Coulomb interactions can be found in Refs.~\onlinecite{nan_lin_prl,Tran16}.

\subsection{GW+DMFT}
GW+DMFT\cite{Sun02,Biermann03,GW_review_werner2016} is based on the premise that both non-local interactions and non-local correlations are important and cannot be discarded; however, the non-local interactions can be treated perturbatively without a significant loss of accuracy. 

The starting point of the GW+DMFT procedure is the GW approximation \cite{Hedin65,Onida02} for which the $\Phi$ functional consist of an infinite series of `bubble' polarization diagrams, $P=GG$, connected by bare interaction lines. This series of bubbles can be resumed into a frequency-dependent `screened' interaction $W=v+vPW$, where $v$ is the bare Coulomb interaction. The self-energy is approximated as $\Sigma=-GW$, so that in the GW approximation $\Phi[G]=-\frac{1}{2}GWG$. 

As \textcite{Albladh99} showed, it is convenient to define a functional $\Psi$, which is a functional both of the Green's function $G$ and of the screened interaction $W$,\cite{Albladh99} as
\begin{align}
\Psi[G,W] = \Phi - \frac{1}{2}(PW-\log(1+PW))
\end{align}
which satisfies
\begin{align}
\left(\frac{\delta\Psi}{\delta W}\right)_G &= -\frac{1}{2}P,\\
\left(\frac{\delta\Psi}{\delta G}\right)_W &= \Sigma.
\end{align}
Together with the Dyson equation that relates $G$ to $\Sigma$, these expression form a closed set of equations that allow the self-consistent computation of $\Sigma$ and $W$. We note that while these equations are $\Phi$ (and $\Psi$)-derivable, and should be solved in a self-consistent manner, the size and complexity of $W$ as well as the difficulty in carrying out the self-consistency necessitates additional approximations~\cite{QPGW_Schilfgaarde,PhysRevB.66.195215,Onida02,GW100} in the case of large realistic systems, which may not respect the conserving properties of Hedin's `fully self-consistent' $GW$ approximation. Notable cases where these equations have been solved self-consistently without any approximations are the electron gas,\cite{Holm98}  atoms and small molecules,\cite{Albladh99,Dahlen04,Stan06,Koval14} and  lattice model systems.\cite{Gukelberger15}

GW+DMFT then makes use of the fact that, given $W$ in all orbitals, there is a natural way of defining an `effective' $W$ in a subset $d$ of correlated orbitals:\cite{Aryasetiawan04} splitting the polarization into a contribution $P_d$ from the `correlated' orbitals and a contribution from all other orbitals, $P=P_d+P_r$, one can define a screened interaction $W_r$ which does not contain any $d$-to-$r$ processes and reformulate $W$ as
\begin{align}
W=[1-W_r P_d]^{-1}W_r,\\
W_r=[1-vP_r]^{-1}v.
\end{align}
This identity is general and independent of the GW approximation. It allows to formulate non-perturbative corrections containing contributions by orbitals exclusively in the correlated subspace $d$ without double counting. Choosing as a subset of orbitals the ones that are local to the unit cell (or, equivalently, a subset of those local to the unit cell), it follows that\cite{Biermann05}
\begin{widetext}
\begin{align}
\Psi_\text{GW+DMFT} &= \Psi^\text{tot}_\text{GW} +\sum_{j=1}^{M} \left([\Psi^I(G_I, W_r)]_j-[\Psi^I_\text{GW}(G_I, W_r)]_j\right).
\end{align}
\end{widetext}
This defines the GW+DMFT approximation to the exact $\Psi$ functional.

The approximation is noteworthy because it is, as it is written, a diagrammatically sound method for solving realistic correlated many-body problems that includes renormalized interactions and non-perturbative local correlations. In practice, numerous technical and theoretical limitations exist. A fully self-consistent solution of the GW problem is technically very challenging. The various approximations employed (quasiparticles, no full self-consistency, etc) at the level of GW along with the difficulty of numerically solving multi-orbital impurity problems with general non-local time-dependent interactions means that the rigorous diagrammatic footing described above is severely approximated in practical implementations of the GW+DMFT method~\cite{Aryasetiawan98}.

\subsection{Comparison of SEET, DMFT, and GW+DMFT}\label{sec:Comparison}
The methods outlined above have several important commonalities. First, they require the self-consistent solution of a $\Phi$ (or $\Psi$)-derivable diagrammatic system. This implies that (provided the equations are actually solved to self-consistency) the important conservation laws are automatically fulfilled. They also consist of two-step procedures: an `outer loop' that entails the solution of a system using a `cheap' method (e.g. GW, GF2, or HF), and an `inner' loop that requires the solution of a quantum impurity problem using non-perturbative techniques. All methods become exact at infinite temperature, at zero interaction, and when the system decouples into separate impurity problems without any inter-impurity interactions.

However, there are several important distinctions between these methods. The first is the choice of correlated orbital space. In DMFT and its variants, correlated subspace orbitals are chosen {\it a priori} to be the local orbitals or a subset of the local orbitals. This was historically motivated by an exact limit of infinite coordination number, \cite{Metzner89,MullerHartmann89} where the self-energy can be shown to reduce to the local form. The locality approximation can be controlled by systematically extending the size of the unit cell in the real\cite{Lichtenstein00,Kotliar01} or reciprocal space,\cite{Hettler98,Maier05} or by introducing diagrammatic expansions in the non-local contributions.\cite{Kusunose06,Toschi07,Rubtsov08,Rubtsov12,Iskakov16}
In contrast, SEET uses insight from a low-order solution of the system to adaptively define the correlated subspace, {\it e.g.} via consideration of the elements of the diagonalized one-body density matrix that are different from 0 or 2. The control parameter used to converge SEET to the exact limit is the size of the correlated subspaces $N^A_i$, which can be systematically increased.

A second major difference between HF+DMFT, GW+DMFT, and SEET(ED-in-GF2 or ED-in-GW) is the way in which non-local interactions are treated. DMFT neglects any contribution from non-local interactions to the  self-energy,  here particularly any contributions from non-local interactions to the local self-energy are neglected. HF+DMFT evaluates the frequency-independent part of the non-local self-energy at the HF level, but any non-local frequency-dependent contribution to the self-energy is neglected, as both interactions and propagators in $\Phi_\text{DMFT}$ are chosen to be local.

\begin{figure}[htb]
  \includegraphics[width=0.3\columnwidth]{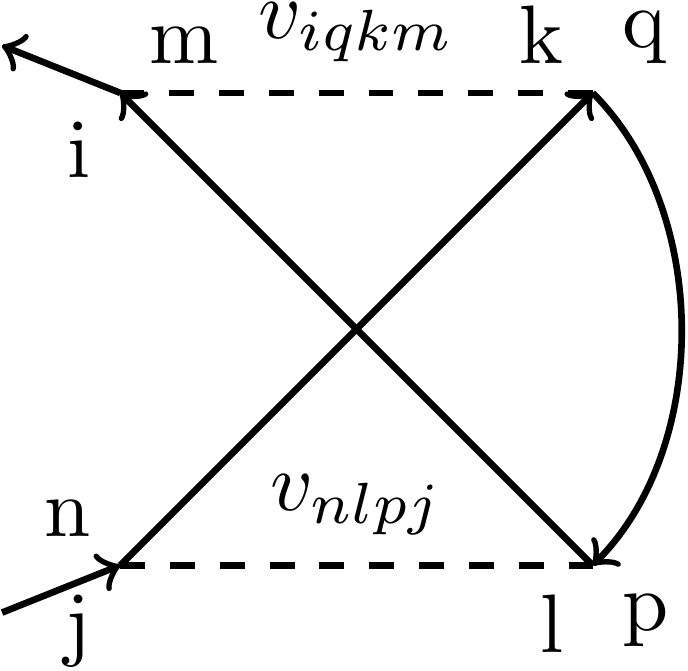} 
  \includegraphics[width=0.3\columnwidth]{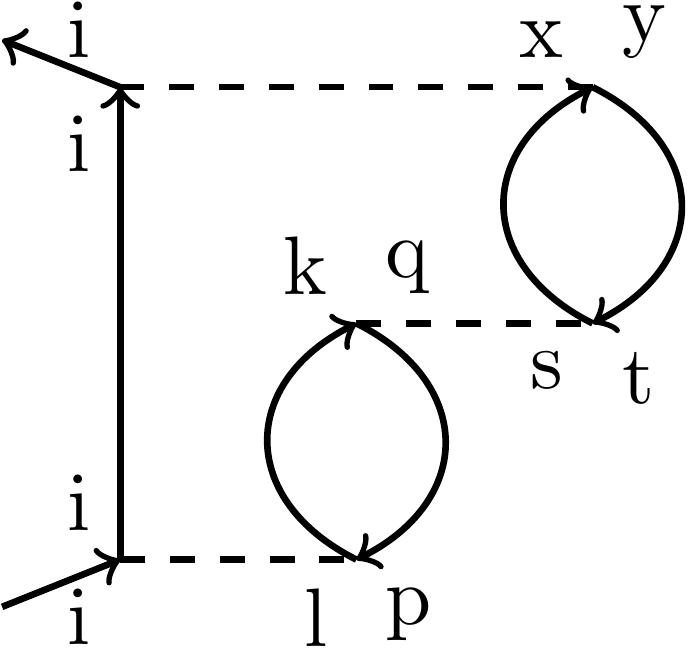} 
  \includegraphics[width=0.22\columnwidth]{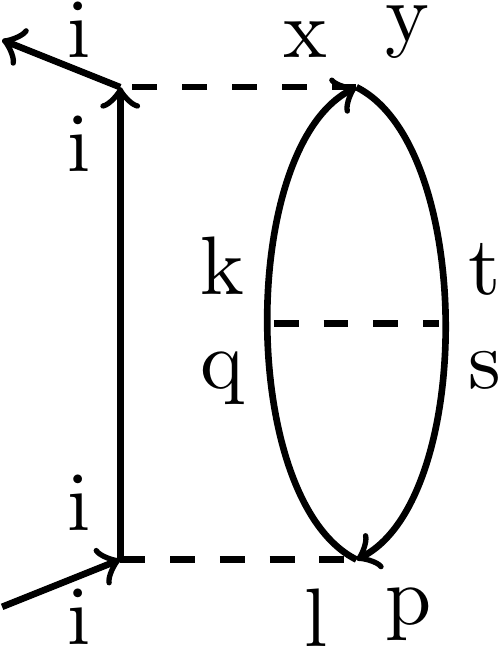} 
  \caption{Left panel: Example of a low order self-energy diagram contained in SEET(ED-in-GF2) but not in GW+DMFT. Middle panel: low order diagram contained in GW+DMFT and SEET(ED-in-GW) but not in SEET(ED-in-GF2). Right panel: low-order diagram not contained in GW+DMFT, SEET(ED-in-GW), and SEET(ED-in-GF2). Dashed lines denote interactions, solid lines Green's functions.}
  \label{fig:lowestSeetNotGW}
\end{figure}
Both GW+DMFT and SEET(ED-in-GF2 or ED-in-GW) include frequency-dependent non-local correlations to some extent. Assuming that a local (rather than an energy) basis is chosen for SEET, the lowest order diagram contained in SEET(ED-in-GF2) but not in GW+DMFT is illustrated in the left panel of Fig.~\ref{fig:lowestSeetNotGW}. 
Here, different indices are assumed to be in different unit cells. Conversely, SEET(ED-in-GF2) in a local basis would not include the diagram illustrated in the middle panel of Fig.~\ref{fig:lowestSeetNotGW}. DMFT could in principle be extended to include the second order exchange diagram, such that the diagram in the left panel is contained, while a formulation of SEET around GW, {\it i.e.} SEET(ED-in-GW), would include the middle panel of Fig.~\ref{fig:lowestSeetNotGW}. None of these methods includes the diagram illustrated in the right panel of Fig.~\ref{fig:lowestSeetNotGW}. As a commonly used basis for SEET is an energy basis, rather than a local basis, a detailed comparison in the practically relevant case is not straightforward.

A third major difference consists of the selection of a basis. As DMFT-type methods perform a local approximation, the choice of basis functions strongly influences the types of correlations that can be contained in DMFT. In contrast, the adaptive choice of SEET basis does not require a localization procedure. 

Finally, the nature of the correlated impurity problem is rather different in SEET and GW+DMFT. GW+DMFT, due to its construction of a screened interaction, requires impurity solvers able to evaluate problems with fully general frequency-dependent interactions. While efficient Monte Carlo methods exist that solve impurity problems with frequency-dependent density-density interactions,\cite{Werner07,Werner10} efficient impurity solvers able to treat general frequency-dependent four-fermion interactions do not yet exist.
SEET, on the other hand, due to the use of the $\Phi$ functional, requires no frequency-dependence in the interactions. However, the rotation to the natural orbital basis in which the density matrix is diagonal usually mixes all orbitals and interactions, necessitating a treatment of the full four-fermion interaction terms (rather than just density-density interactions) with `off-diagonal' hybridization functions.

\section{non-local interactions, correlations, and screening}\label{sec:Screening}
Non-local interactions and non-local dynamical correlations (caused both by local and non-local interactions) alter the  local low-energy physics. A combination of these effects is colloquially summed up under the term `screening', despite very different physical and diagrammatic origins.
As the methods discussed above treat `screening' to a different extent, we briefly discuss various aspects of it.

First, the `screened interaction' $W$ describes a way of re-summing certain classes of diagrams. $W$ then takes the role of the bare interaction $v$ in $\Phi$ and removes diagrams with repeated insertion of polarization parts, at the cost of introducing a frequency dependence.\cite{Hedin65} The need for formulating perturbation theories in powers of $W$ is motivated by a divergence of the perturbation theory in $v$, when truncated at any order, in the infinite system size (momentum $q\rightarrow 0$) limit of the electron gas.\cite{Mahan00} In contrast, a perturbation theory in $W$ removes this divergence and stays finite.\cite{Hedin65} Within GW and GW+DMFT, as well as within SEET(ED-in-GW), terms are included at least to lowest order in $W$, and $W$ is approximated by the lowest order $P$.

SEET(ED-in-GF2) is based on a GF2 starting point that is divergent for metallic systems in the thermodynamic limit, as it is formulated in terms of the bare $v$. However, any finite system will yield a convergent answer. Thus, for a finite system, in an energy basis, the identification of the correlated orbitals will add near-Fermi-surface states to the correlated subspace and converge as the subspace is enlarged. 

A second, entirely different effect also commonly referred to as `screening' that leads to lowering of local bare Coulomb interactions is generated by the effect of non-local interactions on the local self-energy.\cite{Rusakov14}
If the total orbital space is divided into a correlated subspace and the remainder, the correlated subspace self-energy acquires contributions due to non-local interactions with vertices and propagators in the remainder. This effect is  general and present both for the frequency independent and dependent contribution to the self-energy. It is best illustrated for the frequency independent Hartree-Fock contribution $\Sigma_{\infty}$ that can be separated into the following contributions:
\begin{eqnarray}
[\Sigma_{\infty}]_{ij\in A}&=&\sum_{kl}\gamma_{kl}(v_{ikjl}-0.5v_{iklj})\\ \nonumber
&=&[\Sigma_{\infty}]_{ij\in A}^\text{embedded}+[\Sigma_{\infty}]_{ij\in A}^\text{embedding},
\end{eqnarray}
\begin{eqnarray}
[\Sigma_{\infty}]_{ij\in A}^\text{embedded}=\sum_{kl \in A}\gamma_{kl}(v_{ikjl}-0.5v_{iklj}),
\end{eqnarray}
\begin{eqnarray}
[\Sigma_{\infty}]_{ij\in A}^\text{embedding}&=& \sum_{kl \in R} \gamma_{kl} (v_{ikjl}-0.5v_{iklj})\\ \nonumber
&+& \sum_{k \in A}\sum_{l \in R}\gamma_{kl}(v_{ikjl}-0.5v_{iklj}).
\end{eqnarray}
Here the matrix elements $[\Sigma_{\infty}]_{ij\in A}$ have an `embedded' contribution coming only from orbitals belonging to  the subset $A$ and an `embedding' contribution where the summation runs over other orbitals $R$ that are not contained in the subset $A$.

A model with non-local interactions often appears to have a smaller local self-energy than the same model with only on-site interactions.\cite{Ayral13,vanLoon14} Similarly, a multi-orbital model where inter-orbital interactions are truncated to density-density interactions encounters its metal-to-insulator transition at a weaker interaction than one with the full interaction structure.\cite{Werner09,Medici11,Antipov12} As the DMFT approximation neglects all inter-unit-cell interactions inside the correlated subspace, and as technical limitations of the impurity solvers require restriction to density-density terms, the effective DMFT interactions are additionally lowered to account for these corrections. 

In SEET, this method-dependent `screening' contribution that results in the lowering of the correlated orbital subspace self-energy is not caused by introducing effective interactions. Rather, the `embedded' subspace self-energy $[\Sigma^{A}]_{ij\in A}^\text{embedded}=[\Sigma^{A}_\text{strong}]$ is evaluated using the bare Coulomb interactions (transformed to the appropriate basis) and is `screened' due to the presence of the `embedding' self-energy, $[\Sigma]_{ij\in A}^\text{embedding}=[\Sigma^\text{tot}_\text{weak}]_{ij\in A}-[\Sigma^{A}_\text{weak}]$. Note that the internal summations in $[\Sigma^\text{tot}_\text{weak}]_{ij\in A}$ extend over the orbitals that are not present in the correlated subspace, thus accounting for all the effects of the non-local interactions on the total frequency dependent subspace self-energy, $[\Sigma^{A}]=[\Sigma^{A}]_{ij\in A}^\text{embedded}+[\Sigma]_{ij\in A}^\text{embedding}$.
\section{Conclusions}\label{sec:Conclusions}
We have discussed several diagrammatic approximations capable of describing a  full Coulomb Hamiltonian. These approximate methods can then be used in {\em ab initio} calculations of realistic materials or molecular problems. We have paid particular attention to the functional interpretation and have shown that the DMFT - type approximations, where the correlated subspace orbitals are chosen to be local to the unit cell, are a subclass of a wider class of self-energy embedding theories, which can deal with both local and non-local orbitals present in the correlated subspace.

We have also shown that relaxing the locality approximation of the self-energy leads to additional freedom in choosing `correlated' orbitals, and introduces a systematic small parameter that can be controlled in practice. Choosing the correlated orbital subspace as a set of one-body density matrix eigenvectors corresponding to eigenvalues with partial occupancy (most different than 0 or 2) provides an adaptive selection procedure.

While all the methods outlined here have a rigorous theoretical foundation, practical implementations of real-materials embedding calculations remain extremely difficult and the approximations needed to lower the computational cost typically break $\Phi$-derivability.
While some of these approximations have the potential to be removed with future increases of computational power, calculating frequency dependent renormalized interactions in GW+DMFT for impurity models remains challenging.
We therefore believe that embedding methods that do not rely on explicitly renormalized interactions in the correlated subspace, such as SEET(ED-in-GW) and SEET(ED-in-GF2), offer a promising route to the simulation of realistic materials with systematically improvable accuracy.

\begin{acknowledgments}
DZ and EG were supported by the Simons Foundation via the Simons Collaboration on the Many-Electron Problem. We thank Hugo Strand for insightful comments and a careful reading of the manuscript.
\end{acknowledgments}
%

\appendix
\section{Iterative updates in SEET}\label{app:SEETAlgo}
The SEET equations are formulated as a set of self-consistent equations that need to be solved iteratively. The iteration procedure consists of several parts: {\bf(i)} the problem setup with construction of two-body interactions and one-body integrals (hoppings), {\bf(ii)} the solution of the entire system using a low-level, usually weak-coupling method (GF2, GW), {\bf (iii)} the construction of the correlated subspace(s) and impurity problem(s), {\bf (iv)} the solution of the impurity problems using a high-level, usually non-perturbative, impurity solver method, and {\bf(v)} the adjustment of the chemical potential to match the target particle number of the system. 

 The detailed SEET algorithm can be summarized as follows.\\
 {\bf The low level method loop and orbital basis choice.}
 \begin{description}
 \item [IN1] Choose a basis set.
 \item [IN2] In the chosen basis evaluate $t$, and $v$ to represent the Hamiltonian of interest.
 \item [IN3] Solve the Hartree-Fock or Density Functional Theory equations to obtain an initial bare Green's function $G_0$. 
 \item [LL0]  Starting from a given $G_0$, perform self-energy evaluation for the total system with a low level method and obtain $\Sigma^\text{tot}_\text{weak}$, $[\Sigma^\text{A}_\text{weak}]_i$, $\Sigma^\text{R}_\text{weak}$, $\Sigma^\text{int}_\text{weak}$, and $G^\text{tot}_\text{weak}$. 
 \item [LL3]  Choose the most suitable basis for the subsystem $A_i$ that is physically motivated. It can be a energy, occupation, or a local orbital basis.
 \item [LL4]  Transform $t$,  $\Sigma^\text{tot}_\text{weak}$, $[\Sigma^\text{A}_\text{weak}]_i$, $\Sigma^\text{R}_\text{weak}$, $\Sigma^\text{int}_\text{weak}$, and $G^\text{tot}_\text{weak}$ to the new basis. Only a subset of $v$ where all orbital indeces are belonging to subset $A_i$ ( $v_{ijkl} \in A_i$) needs to be transformed to the new basis.
 \end{description}
{\bf The embedding loop.}
 \begin{description}
 \item  [EL1] Construct $G^\text{tot}$ from Eq.~\ref{Gweak} with the self-energy given by Eq.~\ref{eqn:sigma_seet}. 
 Note, that in the first iteration $[\Sigma^\text{A}_\text{strong}]_i=
[\Sigma^\text{A}_\text{weak}]_i$.
\item[EL2] If the particle number is different from the desired particle number, adjust the chemical potential until the desired particle number is reached.
 \item [EL3] Find hybridization $\Delta$ according to Eq.~\ref{eq:hybfun}. 
 \end{description}
 {\bf The high level solver part.}
 \begin{description} [align=right,labelwidth=2cm]
 \item  [HL1] Define a non-interacting impurity Green's function from Eq.~\ref{eqn:g0}. Take care of possible double counting corrections.
 \item [HL2] Evaluate $[\Sigma^\text{A}_\text{strong}]_i$ from Eq.~\ref{eqn:dyson_imp} using a high level solver.
 \end{description}
 \begin{description}
 \item [EL4] Come back to {\bf EL1} and use $[\Sigma^\text{A}_\text{strong}]_i$ from the high level solver. Iterate until $\Delta$ or the electronic energy that can be evaluated at this point will stop to change.
 \item [LL5] Go back to {\bf LL0} and use the resulting $G^\text{tot}$ as that starting Green's function to perform a single update of all the low level self-energies and return to~${\bf EL1}$.
 \end{description}
Note, that in a ``single shot SEET procedure'' the {\bf LL5} point is not executed and the iterative loop is terminated once $\Delta$ or electronic energy evaluated at the {\bf EL4}  point stops to change.

\end{document}